\documentclass{jps-cp}
\usepackage{txfonts} %Please comment out this line unless the txfonts package is availabe in your LaTeX system.
\usepackage{bm}% bold math
\usepackage{amsmath}
\usepackage{color}

\def\gsim{\mathop {\vtop {\ialign {##\crcr 
$\hfil \displaystyle {>}\hfil $\crcr \noalign {\kern1pt \nointerlineskip } 
$\,\sim$ \crcr \noalign {\kern1pt}}}}\limits}
\def\lsim{\mathop {\vtop {\ialign {##\crcr 
$\hfil \displaystyle {<}\hfil $\crcr \noalign {\kern1pt \nointerlineskip } 
$\,\,\sim$ \crcr \noalign {\kern1pt}}}}\limits}
\makeatletter
\newcommand{\subscripts}[3]{%
  \@mathmeasure\z@\displaystyle{#2}%
  \global\setbox\@ne\vbox to\ht\z@{}\dp\@ne\dp\z@
  \setbox\tw@\box\@ne
  \@mathmeasure4\displaystyle{\copy\tw@_{#1}}%
  \@mathmeasure6\displaystyle{{#2}_{#3}}%
  \dimen@-\wd6 \advance\dimen@\wd4 \advance\dimen@\wd\z@
  \hbox to\dimen@{}\mathop{\kern-\dimen@\box4\box6}%
}
\makeatother
\title{Charge Transfer Effect under Odd-Parity Crystalline Electric Field}

\author{Shinji \textsc{Watanabe}$^{1}$ and Kazumasa \textsc{Miyake}$^{2}$}

\inst{$^{1}$Department of Basic Sciences, Kyushu Institute of Technology, Kitakyushu, Fukuoka 804-8550, Japan \\
$^{2}$Center for Advanced High Magnetic Field Science, Osaka University, Toyonaka, Osaka 560-0043, Japan}

\email{swata@mns.kyutech.ac.jp}

\recdate{September 8, 2019}

\abst{
Charge-transfer effect under odd-parity crystalline electric field (CEF) is analyzed theoretically. 
In quantum-critical metal $\beta$-YbAlB$_4$, seven-fold configuration of B atoms surrounding Yb atom breaks local inversion symmetry at the Yb site, giving rise to the odd-parity CEF. 
Analysis of the CEF on the basis of hybridization picture shows that admixture of 4f and 5d wave functions at Yb with pure imaginary coefficient occurs, which makes magnetic-toroidal (MT) and electric-dipole (ED) degrees of freedom active. By constructing the minimal model for periodic crystal $\beta$-YbAlB$_4$, we show that the MT as well as ED fluctuation is divergently enhanced at the quantum critical point of valence transition simultaneously with critical valence fluctuations. 
}

\kword{charge transfer, odd parity CEF, magnetic toroidal fluctuation, electric dipole fluctuation, valence QCP, $\beta$-YbAlB$_4$, $\alpha$-YbAl$_{1-x}$Fe$_x$B$_4$}

\begin{document}
\maketitle

\section{Introduction}
The heavy-electron metal $\beta$-YbAlB$_4$ showing quantum criticality not following the conventional magnetic criticality~\cite{Moriya,Hertz,Millis} has attracted great interest~\cite{Nakatsuji}. 
In particular,  
a new type of scaling called $T/B$ scaling where magnetic susceptibility is expressed as a single scaling function of the ratio of temperature $T$ and magnetic field $B$ was discovered~\cite{Matsumoto}, and the valence of Yb was identified as the intermediate as Yb$^{+2.75}$ at $T=20$~K~\cite{Okawa2009}. 
The quantum criticality as well as the $T/B$ scaling has been shown to be explained by the theory of critical Yb-valence fluctuations~\cite{WM2010,WM2014}. 
Recently, the direct evidence of the valence quantum critical point (QCP) has been observed experimentally. $\alpha$-YbAlB$_4$ is a sister compound of $\beta$-YbAlB$_4$, which shows the Fermi-liquid behavior at low temperatures. However, by substituting Fe into Al 1.4$\%$, the same quantum criticality and $T/B$ scaling as those observed in $\beta$-YbAlB$_4$ appears~\cite{Kuga2018}. At just $x=0.014$ in $\alpha$-YbAl$_{1-x}$Fe$_{x}$B$_4$, a sharp change in Yb valence as well as the volume change has been detected~\cite{Kuga2018}. 
This is the direct experimental verification of the valence QCP with quantum valence criticality proposed theoretically in Refs.~\cite{WM2010,WM2014}. 

The critical valence fluctuations are charge-transfer fluctuations between the 4f electron at Yb and conduction electrons.  
In this paper, we reveal the charge-transfer effect under odd-parity CEF arising from local configuration of atoms around Yb in $\beta$-YbAlB$_4$~\cite{WM2019}.   

First, we analyze the CEF in $\beta$-YbAlB$_4$ on the basis of the hybridization picture taking into account the geometrical symmetry of atoms accurately in Sect.~\ref{sec:CEF}. 
In Sect.~\ref{sec:lattice}, we construct the minimal model for the periodic crystal of $\beta$-YbAlB$_4$ and show that novel multipole degrees of freedom such as the magnetic toroidal (MT) and electric dipole (ED) become active as fluctuations near the QCP of the valence transition. 
The paper is summarized in Sect.~\ref{sec:summary}.

\section{Analysis of CEF on the basis of hybridization picture}
\label{sec:CEF}

In $\beta$-YbAlB$_4$, the CEF ground state was proposed to be the $|J=7/2, J_{z}=\pm 5/2\rangle$ state~\cite{NC2009}, which has been supported by the measurements of 
the anisotropy of the magnetic susceptibility\cite{Nakatsuji},
the M\"{o}ssbauer spectroscopy~\cite{Kobayashi2018} and the linear dichroism in photoemission~\cite{Sekiyama2018}. 
This suggests that the CEF is to be understood from the hybridization picture rather than the point-charge model. This view is compatible with the fact that the conical wave function spreads in the direction almost toward 7 B rings, which acquires the largest 4f-2p hybridization. 
Hence, hereafter we analyze the CEF on the basis of the hybridization picture. 

%%%%%%%%%%%%%% Fig.1 %%%%%%%%%%%%%%%%%%%%%%%%%%%%%%%%%%%%%%%%%
\begin{figure}[tb]
\includegraphics[width=14cm]{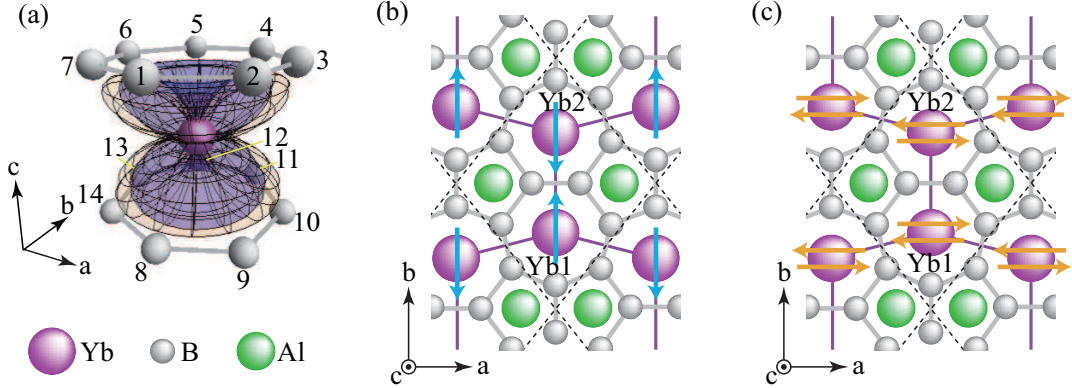}%
\caption{(Color online) (a) 7 B rings at upper and lower planes surround Yb. 
Magnitudes of spherical parts of 4f wave function $\langle\hat{\bm r}|J=7/2, J_z=\pm 5/2\rangle$ (orange) and 5d wave function $\langle\hat{\bm r}|J=5/2, J_z=\pm 3/2\rangle$ (purple) at Yb are plotted. 
(b) Arrow at Yb represents the ED moment. 
(c) Arrows at Yb represent the MT-dipole moments for the Kramers states (see text).
Unit cell is enclosed area by dashed lines in (b) and (c).
}
\label{fig:Yb_B}
\end{figure}
%%%%%%%%%%%%%%%%%%%%%%%%%%%%%%%%%%%%%%%%%%%%%%%%%%%%%%%%%%%%%%

The CEF energy of 4f level is quantified by the 2nd order perturbation as
\begin{eqnarray} 
\Delta E\left(\pm\frac{5}{2},\pm\frac{5}{2}\right)=\subscripts{4f}{\left\langle\pm\frac{5}{2}|\frac{{H_{i1}^{pf}}^{*}H_{i1}^{pf}}{E-H_{0}}|\pm\frac{5}{2}\right\rangle}{4f}\equiv\varepsilon_{f}
\label{eq:Ef}
\end{eqnarray}
with respect to the 4f-2p hybridization
\begin{eqnarray}
H_{i1}^{pf}=
\sum_{\langle{i1,j}\rangle,m,\sigma,J_z}
\left(
V_{jm\sigma,i1J_z}^{pf}p^{\dagger}_{jm\sigma}f_{i1J_z}+{\rm h.c}.
\right). 
\end{eqnarray}
Here, $\langle i1,j\rangle$ denotes the nearest-neighbor (N.N.) B sites and the Yb1 site in the $i$-th unit cell, i.e., $j=1\sim 7$ $(8\sim 14)$ for the upper (lower) plane in Fig.~\ref{fig:Yb_B}(a). 
Here, $m=z, \pm$, $\sigma=\uparrow, \downarrow$, and $J_z=\pm\frac{5}{2}$.

As for Yb 5d state, wave function of the $|J=5/2, J_{z}=\pm 3/2\rangle$ state spreads most closely to 7 B rings among $J=5/2$ and $3/2$ manifolds, which acquires the largest 5d-2p hybridization. 

The CEF energy of the 5d level is quantified by the 2nd order perturbation as 
\begin{eqnarray}
\Delta E\left(\pm\frac{3}{2},\pm\frac{3}{2}\right)=\subscripts{5d}{\left\langle\pm\frac{3}{2}|\frac{{H_{i1}^{pd}}^{*}H_{i1}^{pd}}{E-H_{0}}|\pm\frac{3}{2}\right\rangle}{5d}\equiv\varepsilon_{d} 
\label{eq:Ed}
\end{eqnarray} 
with respect to 5d-2p hybridization
\begin{eqnarray}
H_{i1}^{pd}=
\sum_{\langle{i1,j}\rangle,m,\sigma,J_z}
\left(
V_{jm\sigma,iJ_z}^{pd}p^{\dagger}_{jm\sigma}d_{i1J_z}+{\rm h.c}.
\right), 
\end{eqnarray}
for $J_z=\pm\frac{3}{2}$. 

Interestingly, local inversion symmetry is broken at Yb site owing to surrounding 7 B rings [see Fig.~\ref{fig:Yb_B}(a)]. So, the off-diagonal term is non-vanishing, which is calculated by the 2nd-order perturbation as 
\begin{eqnarray}
\Delta E\left(\pm\frac{3}{2},\pm\frac{5}{2}\right)=\subscripts{5d}{\left\langle\pm\frac{3}{2}|\frac{{H_{i1}^{pd}}^{*}H_{i1}^{pf}}{E-H_{0}}|\pm\frac{5}{2}\right\rangle}{4f}
=
-\sum_{j=1}^{14}\sum_{m\sigma}V_{jm\sigma,i1\pm\frac{3}{2}}^{pd*}
V_{jm\sigma,i1\pm\frac{5}{2}}^{pf}
\frac{
1-n_{jm\sigma}^p
}{\Delta_0+\epsilon_{jm\sigma}^{p}},
\end{eqnarray}
where $\Delta_0 (>0)$ is the excitation energy to the $4f^0$-hole state (i.e., 4$f^{14}$ electron state). Here, $n_{jm\sigma}^{p}$ and $\epsilon_{jm\sigma}^{p}$ are filling and energy of 2p hole, respectively. 
By inputting Slater-Koster parameters~\cite{Slater,Takegahara} and assuming that all 2p hole numbers and energies are the same, we find that the off-diagonal term becomes pure imaginary. 
\begin{eqnarray}
\Delta E\left(\pm\frac{3}{2},\pm\frac{5}{2}\right)=iA,
\end{eqnarray}
where $A$ is a real number. 
Then, by diagonalizing the $2\times 2$ matrix 
\begin{eqnarray}
%\[
\Delta{E_{\pm}}=
%\left(
%\begin{array}{rr}
\begin{bmatrix}
\varepsilon_{f} & -iA \\
iA              & \varepsilon_{d}
\end{bmatrix},  
%\end{array}
%\right),   
%\] 
\end{eqnarray}
%
%for $|\pm\frac{5}{2}\rangle_{4f} \otimes |\pm\frac{3}{2}\rangle_{5d}$,   
we obtain the CEF ground state as the Kramers doublet  
\begin{eqnarray}
|\Psi_{\pm}\rangle=\left(u|\pm\frac{5}{2}\rangle_{4f}+iA|\pm\frac{3}{2}\rangle_{5d}\right)\frac{1}{\sqrt{u^2+A^2}}, 
\label{eq:CEFWF}
\end{eqnarray}
where $u$ is given by $u=[\varepsilon_{f}-\varepsilon_{d}-\sqrt{(\varepsilon_{f}-\varepsilon_{d})^2+4A^2}]/2$.  
An important result here is the admixture of 4f and 5d wave functions occurs with pure imaginary coefficient. 
This implies that odd-parity CEF term 
\begin{eqnarray}
H_{i1}^{\rm opCEF}=iA\left(d_{i1+\frac{3}{2}}^{\dagger}f_{i1+\frac{5}{2}}+d_{i1-\frac{3}{2}}^{\dagger}f_{i1-\frac{5}{2}}\right)+h.c.
\label{eq:CEF_odd}
\end{eqnarray} 
exists at locally inversion symmetry broken Yb site due to surrounding 7 B rings. 
This is the onsite hybridization, which is usually forbidden in centrosymmetric systems. 

The parity mixing in Eq.~(\ref{eq:CEFWF}) makes electric dipole (ED) active at the Yb site. 
%By calculating the expectation values of the ED moment $Q_{i1x}\equiv -ex_{i1}$ and $Q_{i1y}\equiv -ey_{i1}$ with the CEF ground state~Eq.~(\ref{eq:CEFWF}), we obtain~\cite{WM2019} 
%%%%%
%\textcolor{red}
%{
The ED moments defined by $Q_{i1x}\equiv -ex_{i1}$ and $Q_{i1y}\equiv -ey_{i1}$ are expressed in the $|\pm 5/2\rangle_{4f}\otimes |\pm 3/2\rangle_{5d}$ manifold as $Q_{i1\zeta}=Q_{i1\zeta +}+Q_{i1\zeta -}$ with 
\begin{eqnarray}
Q_{i1x+}&=&\frac{e}{\sqrt{5}}d^{\dagger}_{i1+\frac{3}{2}}f_{i1+\frac{5}{2}}+h.c., \  Q_{i1x-}=-\frac{e}{\sqrt{5}}d^{\dagger}_{i1-\frac{3}{2}}f_{i1-\frac{5}{2}}+h.c.,
\\
Q_{1iy+}&=&-i\frac{e}{\sqrt{5}}d^{\dagger}_{i1+\frac{3}{2}}f_{i1+\frac{5}{2}}+h.c., 
\ Q_{1iy-}=-i\frac{e}{\sqrt{5}}d^{\dagger}_{i1-\frac{3}{2}}f_{i1-\frac{5}{2}}+h.c.
\end{eqnarray}
By calculating the expectation values of the ED moment with the CEF ground state~Eq.~(\ref{eq:CEFWF}), we obtain
%}
%%%%%
%
\begin{eqnarray}
\langle\Psi_{\pm}|Q_{i1x}|\Psi_{\pm}\rangle=0, \ \ 
\langle\Psi_{\pm}|Q_{i1y}|\Psi_{\pm}\rangle=
e\frac{10}{7}\sqrt{\frac{3}{10}}
\frac{uA}{u^2+A^2}. 
\end{eqnarray}
This indicates the ED moment exists along the $b$~$(\parallel{y})$ direction, as shown in Fig.~\ref{fig:Yb_B}(b). Since there exist two equivalent Yb atoms in the unit cell where Yb2 site in Fig.~\ref{fig:Yb_B}(b) has opposite sign of $A$ in Eq.~(\ref{eq:CEF_odd}), the ED moment with $-b$ direction exists at the Yb2 site. These states are symmetrically allowed from the viewpoint of the electric-field direction arising from surrounding 7 B rings [see Fig.~\ref{fig:Yb_B}(b)]. 

Recently, novel multipole degree of freedom, magnetic toroidal moment, has been formulated quantum mechanically by Kusunose et al. as  
\begin{eqnarray}
\bm{t}_{l}({\bm r}_i)=\frac{\bm{r}_i}{l+1}\times\left(\frac{2\bm{l}_i}{l+2}+{\boldsymbol \sigma}_{i}\right) 
\label{eq:MT}
\end{eqnarray}
at the site $\bm{r}_i$, 
where $\bm{l}_i$ and ${\boldsymbol \sigma}_i$ are the orbital and spin angular-momentum operators, respectively~\cite{HK2018,HYYK2018}. 
In the $|\pm\frac{5}{2}\rangle_{4f} \otimes |\pm\frac{3}{2}\rangle_{5d}$ manifold, 
%its operators 
the operators of the MT dipole 
are derived as $T_{i1\zeta}=T_{i1\zeta +}+T_{i1\zeta -}$ for $\zeta=x, y$ with 
\begin{eqnarray}
T_{i1x+}&=&-i
\mu_{B}
\frac{15}{14}\sqrt{\frac{3}{10}}
f_{i1+\frac{5}{2}}^{\dagger}d_{i1+\frac{3}{2}}+h.c., \ \  
T_{i1x-}=i
\mu_{B}
\frac{15}{14}\sqrt{\frac{3}{10}}
f_{i1-\frac{5}{2}}^{\dagger}d_{i1-\frac{3}{2}}+h.c.,
\\
T_{i1y+}&=&
-
\mu_{B}
\frac{15}{14}\sqrt{\frac{3}{10}}
f_{i1+\frac{5}{2}}^{\dagger}d_{i1+\frac{3}{2}}+h.c., \ \     
T_{i1y-}=
-
\mu_{B}
\frac{15}{14}\sqrt{\frac{3}{10}}
f_{i1-\frac{5}{2}}^{\dagger}d_{i1-\frac{3}{2}}+h.c.,
\end{eqnarray}
where $\mu_{\rm B}$ is the Bohr magneton. 
For the CEF ground state in Eq.~(\ref{eq:CEFWF}) we obtain 
\begin{eqnarray}
\langle\Psi_{\pm}|T_{i1x}|\Psi_{\pm}\rangle=\mp 
\mu_{B}
\frac{15}{7}\sqrt{\frac{3}{10}}
\frac{uA}{u^2+A^2}, \ \ 
\langle\Psi_{\pm}|T_{i1y}|\Psi_{\pm}\rangle=0. 
\label{eq:T_xy}
\end{eqnarray}
These results indicate that MT dipole moments are aligned along the $a$ direction, which are cancelled each other for the Kramers degenerate states, as shown in Fig.~\ref{fig:Yb_B}(c). 
At the Yb2 site opposite sign of $A$ is realized so that $\langle\Psi_{\pm}|T_{i2x}|\Psi_{\pm}\rangle$  is obtained by converting $\mp$ in the righ hand side (r.h.s.) of Eq.~(\ref{eq:T_xy}) to $\pm$. 
This still makes the net MT-dipole moment zero at the Yb2 site owing to the cancellation by the Kramers-degenerate states.

%-------------------------------------------------------------------------------------
\section{Periodic crystal $\beta$-YbAlB$_4$}
\label{sec:lattice}

\subsection{Model and method}

On the basis of the analyses in Sect.~\ref{sec:CEF},
we construct the minimal model for the low-energy electronic state in the periodic crystal $\beta$-YbAlB$_4$.
The model consists of 4f$J_z=\pm5/2$ and 5d $J_z=\pm3/2$ states at Yb 
and $2p_{\pm}$ and $2p_{z}$ states at B, as follows.  
\begin{eqnarray}
H=\sum_{i\alpha}\left[H_{i\alpha}^{f}+H_{i\alpha}^{pf}+H_{i\alpha}^{pd}
+H^{U_{fd}}_{i\alpha}
\right]
+H^{d}
+H^{p},
\label{eq:EPAM}
\end{eqnarray}
where $\alpha=1,2$ denote the Yb1 and Yb2 sites, respectively [see Fig.~\ref{fig:Yb_B}(b)]. 
The 4f-hole term is  
\begin{eqnarray}
H_{i\alpha}^{f}=\varepsilon_{f}\sum_{J_z=\pm5/2}n^{f}_{i\alpha J_z}
+Un_{i\alpha +5/2}^{f}n_{i\alpha -5/2}^{f}, 
\end{eqnarray}
where $\varepsilon_{\rm f}$ is the 4f-hole level and $U$ is the onsite Coulomb repulsion between 4f holes. 
The 5d-hole term is
\begin{eqnarray}
H^{d}=\varepsilon_{d}\sum_{i}\sum_{\alpha=1,2}\sum_{J_z=\pm 3/2}n^{d}_{i\alpha J_z}
%\nonumber
%\\
+\sum_{\langle{i\alpha,i'\alpha'}\rangle}\sum_{J_z,J_z'=\pm 3/2}t_{i\alpha J_z,i'\alpha'J_{z}'}^{dd}d^{\dagger}_{i\alpha J_z}d_{i'\alpha'J_{z}'}, 
\end{eqnarray}
where $\varepsilon_{d}$ is the 5d-hole level and $t_{i\alpha J_z,i'\alpha'J_{z}'}^{dd}$ is the transfer of 5d holes between Yb sites with $\langle{i\alpha,i'\alpha'}\rangle$ being taken for the N.N. Yb pairs. 
The onsite Coulomb repulsion between 4f and 5d holes at Yb is 
\begin{eqnarray}
H_{i\alpha}^{U_{fd}}=U_{fd}\sum_{J_z=\pm5/2}\sum_{J_z'=\pm 3/2}n_{i\alpha J_z}^{f}n_{i\alpha J_z'}^{d}.
\end{eqnarray}
The 2p-hole term is 
\begin{eqnarray}
H^{p}=\sum_{\langle{j,j'}\rangle\sigma}\sum_{m,m'=z,\pm}t_{jm,j'm'}^{pp}p^{\dagger}_{jm\sigma}p_{j'm'\sigma},
\end{eqnarray}
where $t_{jm,j'm'}^{pp}$ is the transfer of 2p holes between B sites with $\langle{j,j'}\rangle$ being taken for the N.N. B pairs in the $ab$ plane and in the $c$ direction [see Figs.~\ref{fig:Yb_B}(a) and \ref{fig:Yb_B}(b)]. 
Here we note that the odd-parity CEF term in Eq.~(\ref{eq:CEF_odd}) is {\it not} included explicitly, because it inheres via the 4f-2p and 5d-2p hybridizations.

To clarify the charge-transfer effect under odd-parity CEF, we apply the slave-boson mean-field theory for $U=\infty$~\cite{Read,OM2000} to the Hamiltonian Eq.~(\ref{eq:EPAM}). 
As for the Slater-Koster parameters, 
following the argument of the linear muffin-tin orbital (LMTO)~\cite{Andersen1}, 
we set 
$(pp\pi)=-(pp\sigma)/2$, $(pd\pi)=-(pd\sigma)/\sqrt{3}$, $(pf\pi)=-(pf\sigma)/\sqrt{3}$, $(dd\pi)=-2(dd\sigma)/3$, and $(dd\delta)=(dd\sigma)/6$. 
Then, we set $(pp\sigma)=-1.0$ in the hole picture, which is taken as the energy unit hereafter, and set   
$(pd\sigma)=0.6$, $(pf\sigma)=-0.3$, $(dd\sigma)=0.4$, and 
$\varepsilon_{d}=-1.0$ at filling $\bar{n}=23/32$  
so as to reproduce the recently-observed energy band near the Fermi level by the photoemission measurement~\cite{Cedric}. 
Here, the filling is defined by $\bar{n}\equiv\sum_{\alpha=1,2}(\bar{n}^{f}_{\alpha}+\bar{n}^{d}_{\alpha})/4+\sum_{j=1}^{8}\bar{n}^{p}_{j}/16$, 
where $\bar{n}^{\eta}_{\alpha}$  for $\eta=f$, $d$ and $\bar{n}^{p}_{j}$ are defined by 
$\bar{n}^{\eta}_{\alpha}=\sum_{iJ_z}\langle n^{\eta}_{i\alpha J_z}\rangle/N$ and   
$\bar{n}^{p}_{j}\equiv\sum_{{\bm k}m\sigma}\langle p^{\dagger}_{{\bm k}jm\sigma}p_{{\bm k}jm\sigma}\rangle/(3N)$ respectively with $N$ being the number of unit cells. 

By solving the mean-field equations, 
we obtained the paramagnetic ground state for various $\varepsilon_f$ and $U_{fd}$ in the $N=8^3, 16^3$ and $32^3$ systems (see Ref.~\cite{WM2019} for details). 
The results in $N=32^3$ will be shown below. 

\subsection{Results}

Figure~\ref{fig:nfnd}(a) shows 
the $\varepsilon_{f}$ dependence of $\bar{n}_{f}($=$\bar{n}^f_1$=$\bar{n}^f_2)$ for $U_{fd}=
1.00$, $1.20$, $1.35$, $1.40$, and $1.45$. 
As $U_{fd}$ increases, $\bar{n}^f$ change becomes steeper and the slope diverges at $(\varepsilon_f^{\rm QCP},U_{fd}^{\rm QCP})\approx(-2.3001, 1.40)$, indicating diverging critical valence fluctuations at the valence QCP $\chi_v\equiv -\partial\bar{n}_f/\partial\varepsilon_{f}=\infty$. 
As $U_{fd}$ exceeds $U_{fd}^{\rm QCP}$, a jump in $\bar{n}_f$ appears, indicating the first-order valence transition. 

Our realistic minimal model shows the valence QCP with $\bar{n}_{f}=0.71$ corresponding to Yb$^{+2.71}$, which is consistent with the intermediate valence of Yb observed in $\beta$-YbAlB$_4$ (Yb$^{+2.75}$ at $T=20$~K)~\cite{Okawa2009} and $\alpha$-YbAl$_{0.086}$Fe$_{0.014}$B$_4$ (Yb$^{+2.765(5)}$ at $T=20$~K)~\cite{Kuga2018}. 
The value of the 4f-5d Coulomb repulsion is evaluated to be $U_{fd}^{\rm QCP}\approx 6.6$~eV,  
if we employ $(pp\sigma)\approx 4.7$~eV as a typical value taken from the result of first-principles band-structure calculation for B~\cite{Papa}. 
Since $U_{fd}$ is onsite interaction at the Yb site, this value seems reasonable. 
To examine this point, recently-developed partial-fluorescence-yield (PFY) measurement of X ray using Yb $L_3$ edge~\cite{Tonai2017} seems useful for direct evaluation of $U_{fd}$ in $\beta$-YbAlB$_4$ and $\alpha$-YbAl$_{0.086}$Fe$_{0.014}$B$_4$. Such measurements are interesting future subjects.

%%%%%%%%%%%%%% Fig.2 %%%%%%%%%%%%%%%%%%%%%%%%%%%%%%%%%%%%%%%%%
\begin{figure}[tb]
\includegraphics[width=14cm]{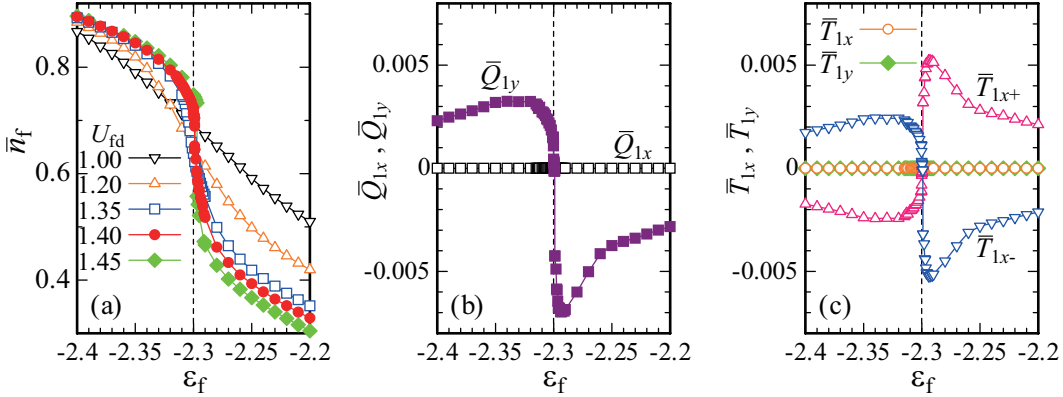}%
\caption{(color online) 
(a) The $\varepsilon_f$ dependence of the 4f-hole number $\bar{n}_f$ at Yb for $U_{fd}=
1.00$, $1.20$, $1.35$, $1.40$, and $1.45$. 
The $\varepsilon_f$ dependences of (b) ED moment and (c) MT-dipole moment at the Yb1 site for 
$U_{fd}=1.40$. In the plot of the  ED (MT-dipole) moment, we set $e=1$ $(\mu_{B}=1)$. 
In (b) and (c), dashed lines denote $\varepsilon_{f}=\varepsilon_{f}^{\rm QCP}$. 
}
\label{fig:nfnd}
\end{figure}
%%%%%%%%%%%%%%%%%%%%%%%%%%%%%%%%%%%%%%%%%%%%%%%%%%%%%%%%%%%%%%

%%%%%
%\textcolor{blue}
%{
As for the Yb2 site, we confirmed $\bar{Q}_{2y}=-\bar{Q}_{1y}$ and $\bar{Q}_{2x}=0$ as well as $\bar{T}_{2x}=0$ and $\bar{T}_{2y}=0$, which are consistent with the second-order perturbation analysis in Sect.2.
%}
%%%%%
 
Figure~\ref{fig:nfnd}(b) shows the 4f-level dependences of the ED moment. Here, $\bar{X}_{\alpha\zeta}$ for $X=Q, T$ and $\zeta=x, y$ is defined by  
$\bar{X}_{\alpha\zeta}\equiv \bar{X}_{\alpha\zeta+}+ \bar{X}_{\alpha\zeta-}$ with $\bar{X}_{\alpha\zeta\pm}=\frac{1}{N}\sum_{i}\langle X_{i\alpha\zeta\pm}\rangle$. 
The results of $\bar{Q}_{1x}=0$ and $\bar{Q}_{1y}\ne 0$ indicate that 
the ED moment along $b$ direction exists. 
The $b$ component of the ED moment 
$\bar{Q}_{1y}$ shows a sign change at the valence QCP indicated by the vertical dashed line in Fig.~\ref{fig:nfnd}(b). 

Figure~\ref{fig:nfnd}(c) shows the 4f-level dependences of the MT-dipole moment. 
The results of $\bar{T}_{1y}=0$ and $\bar{T}_{1x+}\ne 0$ indicate that 
the MT-dipole moment along $a$ direction exists, which also shows the sign change at valence QCP. On the other hand, the MT-dipole moment $\bar{T}_{1x-}$ for another state of the Kramers pair has opposite sign. 
Hence, the net MT-dipole moment is zero $\bar{T}_{1x}=\bar{T}_{1x+}+\bar{T}_{1x-}=0$. 
Here we note that the ED and MT-dipole moments satisfy the following relation 
\begin{eqnarray}
\bar{T}_{1x-}=\bar{T}_{1x+}=\frac{3\mu_{\rm B}}{4e}\bar{Q}_{1y}, 
\end{eqnarray}
which confirms the general formulation of multipoles in Ref.~\cite{HK2018}. 
These results are consistent with the second-order-perturbation analysis discussed in Sect.~\ref{sec:CEF}, which indicates that the odd-parity CEF in Eq.~(\ref{eq:CEF_odd}) certainly inheres in the Hamiltonian Eq.~(\ref{eq:EPAM}). 
As noted in Sect.~\ref{sec:CEF}, the Yb2 site in Fig.~\ref{fig:Yb_B}(b) has opposite sign of $A$ in Eq.~(\ref{eq:CEF_odd}), and hence the ED and MT-dipole moments at the Yb2 site direct oppositely to those in the Yb1 site, as shown in Figs.~\ref{fig:Yb_B}(b) and \ref{fig:Yb_B}(c). 
We note that the results in Fig.~\ref{fig:nfnd}(b) show that recovery of local inversion symmetry occurs at the valence QCP, where the sign changes of the ED and MT-dipole occurs. 

\begin{figure}[tb]
\includegraphics[width=16cm]{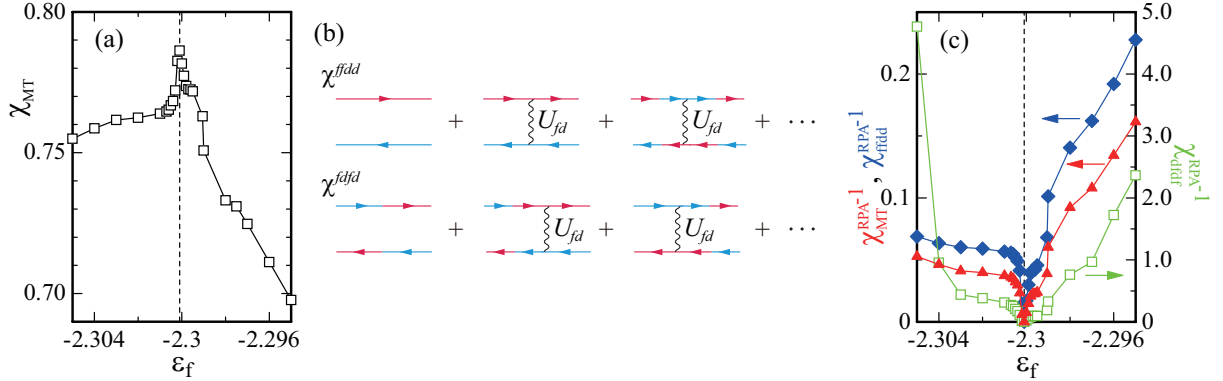}%
\caption{(Color online) (a) The $\varepsilon_{f}$ dependence of $\chi_{\rm MT}$ calculated by the mean-field theory for $U_{fd}=U_{fd}^{\rm QCP}$. 
(b) Diagrams for RPA formalism with respect to $U_{fd}$ for $\chi^{ffdd}_{\nu\nu}({\bm q},\omega)$ 
and $\chi^{fdfd}_{\nu\nu}({\bm q},\omega)$.
Red (light blue) arrow represents the Green function of 4f (5d) quasiparticle and 
wevy line represents $U_{fd}$.
(c) The $\varepsilon_f$ dependences of ${\chi_{MT}^{\rm RPA}}^{-1}$ 
(filled triangle), 
${\chi_{ffdd}^{\rm RPA}}^{-1}$ (filled diamond), and ${\chi_{dfdf}^{\rm RPA}}^{-1}$ 
(square) calculated by RPA for $U_{fd}=U_{fd}^{\rm c}$ (see text).  
In (a) and (c), the dashed line denotes $\varepsilon_{f}=\varepsilon_{f}^{\rm QCP}$ and we set $\mu_{\rm B}=1$. 
}
\label{fig:MT_EfA}
\end{figure}

Although vanishing MT-dipole moment is a consequence of the time-reversal symmetry of the paramagnetic state, its fluctuation can arise even in the paramagnetic state. 
Actually, we find that the MT-dipole susceptibility $\chi_{\rm MT}$ has a peak at the valence QCP as shown in Fig.~\ref{fig:MT_EfA}(a). 
Here, MT-dipole susceptibility $\chi_{\rm MT}\equiv\lim_{{\bm q}\to{\bm 0}}\chi_{T_{x}T_{x}}({\bm q},0)$ is defined as
\begin{eqnarray}
\chi_{T_{x}T_{x}}({\bm q},\omega)
=\frac{i}{N}\int_{0}^{\infty}dt e^{i\omega t}
\langle[T^{x}_{\bm q}(t), T^{x}_{-\bm q}(0)]\rangle
\end{eqnarray}
with $T_{\bm q}^{x}=\sum_{i}e^{-i{\bm q}\cdot{{\bm r}_{i}}}T_{i1x}$. 
This indicates that the MT fluctuation is enhanced at the valence QCP.

%%%%%%%%%%%%%% Fig.3 %%%%%%%%%%%%%%%%%%%%%%%%%%%%%%%%%%%%%%%%%
%%%%%%%%%%%%%%%%%%%%%%%%%%%%%%%%%%%%%%%%%%%%%%%%%%%%%%%%%%%%%%

\subsection{RPA beyond the mean-field theory}
\label{sec:RPA}

Since the charge-transfer fluctuation diverges at the valence QCP~\cite{WM2010}, the MT fluctuation is expected to diverge at the QCP, if the effect of critical valence fluctuations are taken into account.  
To quantify this beyond the mean-field theory, we rewrite 
the MT susceptibility as the linear combination of the charge-transfer-type susceptibility as follows:  
%
%$
\begin{eqnarray}
\chi_{T_{x}T_{x}}({\bm q},\omega)
%%%%%
%\textcolor{blue}
\approx
%%%%%
\frac{9\mu_{B}^2}{28}
\sum_{\nu=\pm}\left[\chi_{\nu\nu}^{ffdd}({\bm q},\omega)+\chi_{\nu\nu}^{ddff}({\bm q},\omega)
-\chi_{\nu\nu}^{dfdf}({\bm q},\omega)-\chi_{\nu\nu}^{fdfd}({\bm q},\omega)\right],
\label{eq:chiTxTx}
\end{eqnarray}
%$
%
where $\chi_{\nu\nu}^{\beta\gamma\delta\eta}({\bm q},\omega)$ is defined by 
\begin{eqnarray}
\chi_{\nu\nu}^{\beta\gamma\delta\eta}({\bm q},\omega)\equiv\frac{i}{N}\int_{0}^{\infty}dt
e^{i\omega t}\langle[\Delta_{{\bm q}\nu}^{\gamma\delta}(t), \Delta_{{-\bm q}\nu}^{\eta\beta}(0)]\rangle
\label{eq:chi_CT}
\end{eqnarray}
with 
$
\Delta_{{\bm q}\pm}^{fd}=\sum_{\bm k}
f^{\dagger}_{{\bm k}+{\bm q}1\pm\frac{5}{2}}d_{{\bm k}1\pm\frac{3}{2}}
$
and 
$
\Delta_{{\bm q}\pm}^{df}=\sum_{\bm k}
d^{\dagger}_{{\bm k}+{\bm q}1\pm\frac{3}{2}}f_{{\bm k}1\pm\frac{5}{2}}
$. 
%%%%%
%\textcolor{red}
%{
Here we note that the off-diagonal Green function with respect to the Kramers states, which appears via the hybridization between 4f and conduction holes, is negligible because the hybridization is strongly suppressed in the heavy-fermion system $\beta$-YbAlB$_4$, which gives the reduction of the order of $10^{-2}\sim 10^{-3}$. 
%}
%%%%%
%%%%%
%\textcolor{blue}
%{
Within this approximation, the right hand side of Eq.~(\ref{eq:chiTxTx}) is expressed as above. 
%}
%%%%%

Near the valence QCP, charge-transfer fluctuations are enhanced, which are caused by inter-orbital Coulomb repulsion $U_{fd}$. These effects can be taken into account by the Random Phase Approximation (RPA) with respect to $U_{fd}$ as the corrections to the mean-field state, as shown in Fig.~\ref{fig:MT_EfA}(b).
\begin{eqnarray}
\hat{\chi}({\bm q},\omega)=\hat{\chi}_{0}({\bm q},\omega)[\hat{1}-\hat{U}\hat{\chi}_{0}({\bm q},\omega)]^{-1},
\label{eq:RPA}
\end{eqnarray}
where $\hat{\chi}$, $\hat{\chi}_0$, and $\hat{U}$ are given by
\begin{equation}
\hat{\chi}_{(0)}({\bm q},\omega)=
\begin{bmatrix}
\chi^{ffdd}_{(0)}({\bm q},\omega) & \chi^{fdfd}_{(0)}({\bm q},\omega) \\
\chi^{dfdf}_{(0)}({\bm q},\omega) & \chi^{ddff}_{(0)}({\bm q},\omega)
\end{bmatrix},
\hat{U}=
\begin{bmatrix}
U_{fd} & 0 \\
0 & U_{fd}
\end{bmatrix},
\nonumber
\end{equation}
and $\hat{1}$ is the identity matrix. 
Here, the index 0 specifies the susceptibility calculated for the mean-field state and we omit the index $\nu$ since $\chi^{\beta\gamma\delta\eta}_{++}=\chi^{\beta\gamma\delta\eta}_{--}$ holds in the paramagnetic state. 

%%%%%
%\textcolor{red}
%{
Here we note the ground of Eq. (\ref{eq:RPA}) as follows: The bubble-type diagrams which also appear in the RPA formalism give negligible contributions since these types consist of the Green functions via the strongly-renormalized hybridizations between f and conduction holes of the order of $10^{-2}\sim 10^{-3}$ in the heavy-fermion systems. In Ref. [7], the authors showed that the charge-transfer susceptibilities expressed as the ladder-type diagrams in Fig.3(b) which constitute of the valence susceptibility are derived from the extended periodic Anderson model. By taking into account the effects of the mode-mode coupling of the charge-transfer fluctuations in Fig.3(b), it was shown that the quantum valence criticality emerges, which explains coherently all the measured non-Fermi-liquid behaviors in the physical quantities in $\beta$-YbAlB$_4$~\cite{WM2010,WM2014}. In that paper, we discuss the divergence of the charge-transfer fluctuations shown in Fig.3(b) at the valence QCP within the RPA level. Since the two Yb sites in the unit-cell [see Fig.1(b)] are equivalent crystallographically and we discuss the paramagnetic state without applying the magnetic field, the RPA susceptibility only for the Yb1 site is shown here. For the basis of the RPA formalism with respect to the Coulomb repulsion between f and conduction electrons  near the valence QCP in heavy-fermion systems, we refer to Ref.~\cite{M2007} for readers.
%}
%%%%%

Within this RPA formalism, the critical point is obtained by solving ${\rm det}\{\hat{1}-\hat{U}\hat{\chi}_{0}({\bm q},\omega)\}=0$, which gives $U_{fd}^{\rm c}=1.635$ and $\varepsilon_{f}=\varepsilon_{f}^{\rm QCP}$. 
Then, by substituting $U_{fd}=U_{fc}^{\rm c}$ and $\varepsilon_{f}$ into Eq.~(\ref{eq:RPA}) with $\chi^{\beta\gamma\delta\eta}_{(0)}({\bm q},\omega)$ for $\beta\gamma\delta\eta=ffdd$ and $fdfd$, where $\chi^{ddff}_{(0)}({\bm q},\omega)=\chi^{ffdd}_{(0)}({\bm q},\omega)$ and $\chi^{fdfd}_{(0)}({\bm q},\omega)=\chi^{dfdf}_{(0)}({\bm q},\omega)$ hold, we obtain corresponding $\chi^{\beta\gamma\delta\eta}({\bm q},\omega)$ in the RPA. 
By substituting the obtained $\chi^{\beta\gamma\delta\eta}({\bm q},\omega)$ into the r.h.s. of Eq.~(\ref{eq:chiTxTx}), we obtain the MT-dipole susceptibility in the RPA. 
Figure~\ref{fig:MT_EfA}(c) shows the 4f-level dependence of the RPA susceptibilities of the charge transfers 
$\chi_{ffdd}^{\rm RPA}\equiv\lim_{{\bm q}\to{\bm 0}}\chi^{ffdd}({\bm q},0)$, $\chi_{dfdf}^{\rm RPA}\equiv\lim_{{\bm q}\to{\bm 0}}\chi^{dfdf}({\bm q},0)$, 
and the MT-dipole $\chi_{MT}^{\rm RPA}\equiv\lim_{{\bm q}\to{\bm 0}}\chi_{T_{x}T_{x}}({\bm q},0)$. 
These results show that the 
charge-transfer fluctuations diverge at the valence QCP $\chi_{ffdd}^{\rm RPA}=\infty$ and $\chi_{dfdf}^{\rm RPA}=\infty$, giving rise to the divergence of MT fluctuation $\chi_{MT}^{\rm RPA}=\infty$. Namely, at the valence QCP, critical valence fluctuations diverge. At the same time, the MT-dipole fluctuation diverges 
%%%%%
%\textcolor{blue}
%{
within the present RPA formalism. 
%}
%%%%%

Since the ED susceptibility is defined as  
\begin{eqnarray}
\chi_{Q_{y}Q_{y}}({\bm q},\omega)
=\frac{i}{N}\int_{0}^{\infty}dt e^{i\omega t}
\langle[Q^{y}_{\bm q}(t), Q^{y}_{-\bm q}(0)]\rangle
\end{eqnarray}
with $Q_{\bm q}^{y}=\sum_{i}e^{-i{\bm q}\cdot{{\bm r}_{i}}}Q_{i1y}$, 
the ED susceptibility is proportional to the MT-dipole susceptibility as  
\begin{eqnarray}
\chi_{Q_{y}Q_{y}}({\bm q},\omega)
%%%%%
%\textcolor{blue}
\approx
%%%%%
\frac{4e^2}{9\mu_{B}^2}
\chi_{T_{x}T_{x}}({\bm q},\omega).  
\label{eq:Q_T}
\end{eqnarray}
%
%%%%%
%\textcolor{blue}
%{
This relation is obtained within the approximations where the Green functions for the off-diagonal states with respect to the Kramers state are neglected as noted below Eq.~(\ref{eq:chi_CT}).
%}
%%%%%
Namely, the ED fluctuation along the $b$ direction is proportional to the MT fluctuation along the $a$ direction [see Figs.~\ref{fig:Yb_B}(b) and \ref{fig:Yb_B}(c)].
Hence, 
%it turns out 
%%%%%
%\textcolor{blue}
Eq.~(\ref{eq:Q_T}) suggests
%%%%%
that ED fluctuation also diverges at the valence QCP. This implies that measurement of the ED fluctuation can detect the MT-dipole fluctuation. 

\subsection{Discussions}

We confirmed divergence of the MT-dipole fluctuation also occurs $\partial \bar{T}_{\alpha x}/\partial h_{\alpha}=\infty$ within the mean-field theory by applying conjugate field $-\sum_{i\alpha}h_{\alpha}T_{i\alpha x}$ to Eq.~(\ref{eq:EPAM}). 
Hence, 
%divergence
%%%%%
%\textcolor{blue}
divergent enhancement
%%%%%
 of the MT as well as ED fluctuation at the valence QCP 
%occurs irrespective of theoretical schemes. 
%%%%%
%\textcolor{red}
%{
is expected to occur irrespective of theoretical schemes which describe the divergence of the valence susceptibility at the valence QCP.
%}
%%%%%

%%%%%
%\textcolor{blue}
%{
Since the Yb1 site and Yb2 site are located each other at the inversion-symmetrical positions, we expect that the enhancement of the uniform susceptibility at ${\bm q}={\bf 0}$ of the ED $Q_{1y}$ at the Yb1 site shown above yields the enhancement of the antiferro (AF)-type ED fluctuation with respect to the Yb2 site in the unit cell as shown in Fig.~\ref{fig:Yb_B}(b). As for the MT dipole moment, we also expect that the AF-type fluctuation between the Yb1 and Yb2 sites in the unit cell is enhanced. To confirm this explicitly, it is necessary to calculate the RPA susceptibility in Eq.~(\ref{eq:RPA}) taking into account the two Yb sites. Furthermore, as noted below Eq.~(\ref{eq:chi_CT}), we neglected the off-diagonal component with respect to the Kramers state in the Green function in deriving Eq.~(\ref{eq:chiTxTx}) and also Eq.~(\ref{eq:Q_T}). Within these approximations, we have shown that the MT as well as ED fluctuation diverges at the valence QCP. Although our result suggests that the MT as well as ED fluctuation is at least divergently enhanced near the valence QCP, it is necessary to confirm which fluctuation among the ferro-type or AF type MT and ED fluctuations in the unit cell really diverge simultaneously with the CVF beyond the above-mentioned approximations. To check this point, it is necessary to take into account the contributions from the off-diagonal components with respect to the Kramers states as well as the two Yb sites in the unit cell in Eq.~(\ref{eq:RPA}). Such a calculation is left for the future subject.
%}
%%%%%

Although we focused on the ED and MT-dipole degrees of freedom in this paper,  
%divergence 
%%%%%
%\textcolor{blue}
enhancement
%%%%%
of fluctuations of multipoles expressed by linear combinations of charge-transfer type operators is expected to occur generally at valence QCP.
Since $\pm 1$ different magnetic quantum number from those of the spherical harmonics constituting the 4f wave function $\varphi^{4f}_{\pm}(\hat{\bm r})\equiv\langle\hat{\bm r}|J=7/2, J_z=\pm 5/2\rangle$ of the CEF ground state is necessary to have finite matrix elements of the MT and ED,   
Yb 5d state which consists of the spherical harmonics $Y_{2,\pm 2}(\hat{\bm r})$ and/or $Y_{2,\pm 1}(\hat{\bm r})$ makes the MT and ED moments active~\cite{WM2019}. 
Hence, our results are considered to be relevant to the case that either or some of these four states among 5 orbitals for 5d electron contribute to the conduction band near the Fermi level. 
Detection of MT and ED fluctuations by various experimental probes as noted in Sect.~\ref{sec:RPA} are interesting future subjects.

\section{Summary}
\label{sec:summary}

In $\beta$-YbAlB$_4$, seven-fold configuration of B atoms surrounding the Yb atom breaks local inversion symmetry at the Yb site. On the basis of the hybridization picture, we have shown that admixture of 4f and 5d wave functions with pure imaginary coefficient occurs, giving rise to ED and MT degrees of freedom. By constructing the realistic minimal model, we have shown that at the valence QCP, MT as well as ED fluctuation 
%diverges 
%%%%%
%\textcolor{blue}
is divergently enhanced
%%%%%
simultaneously with critical valence fluctuations. Our study has revealed the underling mechanism by which novel multipole degrees of freedom can be active as fluctuations, which is a new aspect of the charge-transfer effect.

\section*{Acknowledgments}
We acknowledge H. Kusunose, S. Hayami, M. Yatsushiro, H. Kobayashi, C. Bareille, M. Suzuki and H. Harima for valuable discussions. 
This work was supported by JSPS KAKENHI Grant Numbers JP18K03542, JP18H04326, and JP17K05555.

%--------------------------------------------------------------------------------------------

\end{document}